# One Hundred Years of Observations of the Be Star HDE 245770 (the X-ray Binary A0535+26/V725 Tau): The End of an Active Phase?

V. M. Lyuty* and G. V. Zaĭtseva

*Sternberg Astronomical Institute, Universitetskiĭ pr. 13, Moscow, 119899 Russia*
Received April 29, 1999; in final form, July 20, 1999

**Abstract**—*UBV* observations of the X-ray binary system A0535+26/V725 Tau at the Crimean Station of the Sternberg Astronomical Institute in 1980–1998 are presented. Based on our and published data, we analyze the photometric history of the star from 1898. Over a period of 100 years, the star apparently showed all three activity phases (B, Be, Be-shell) of Be stars. We conclude that the X-ray activity of the object is attributable to the 1970–1997 outburst of the Be star due to envelope ejection. The star's colors during the minimum light of 1998 and its 1953–1956 colors (before the outburst) correspond to the spectral type B0–B1 at the color excesses $E_{B-V} = 0.74$ and $E_{U-B} = 0.48$, in agreement with the current spectral type O9.7. The minimum light of 1998 and the color excesses are used to determine the colors of the additional radiation, analyze their evolution during the 1973–1997 outburst, and refine the distance to the object (3 kpc). The colors of the additional radiation at maximum light of the star (1973–1980) match the colors of a hydrogen plasma with $T_e = 1.5 \times 10^4$ K which is optically thick in the Balmer continuum. The brightness decline corresponds to a decrease in the optical depth of the plasma; at $V \simeq 9^m\!.1$, it becomes optically thin in the Balmer continuum with $T_e = 10^4$ K and $N_e = 10^{10}$–$10^{12}$ cm$^{-3}$. This conclusion is consistent with the model of a circumstellar envelope but is inconsistent with the existence of an accretion disk around the neutron star. All the additional radiation responsible for the optical variability is produced by a single source. The intensity of the Hα emission line at maximum light (1975–1980) is triple its intensity in 1987–1997, when quasi-periodic light fluctuations with $P \approx 1400^d$ were observed [1]. At this time, the line intensity correlated with brightness. The Hα line was in absorption at the minimum of 1998, and, at present, the star's active phase appears to have ended. © *2000 MAIK "Nauka/Interperiodica"*.

## INTRODUCTION

The transient X-ray source A0535+26, which was identified with a variable Be star that was subsequently designated as V725 Tau, has been extensively studied since April 1975, when the first of the so-called "giant" X-ray outbursts was recorded. This is believed to be a binary system with a neutron star, a pulsar with a period of 103 s and an orbital period of $111^d$ (see the review article of Giovannelli and Graziati [2]). The orbit is highly eccentric, and X-ray outbursts are observed during periastron passage. The object belongs to massive X-ray binaries in which the primary is a star of spectral type O9e–B1e, luminosity class III–V, and mass 10–20$M_\odot$, while the secondary is a neutron star. In particular, the spectral type of the optical star in A0535+26/V725 Tau was estimated to be O9.7 IIIe [3]. The material that accretes onto the neutron star is mainly supplied through mass loss from a rapidly rotating Be star and stellar wind. Such systems are distinguished by long periods (up to several hundred days) and high orbital eccentricities, which appears to produce the phenomenon of a transient, flaring X-ray source (an X-ray outburst is observed during periastron passage). In contrast to the "standard" massive close (Cyg X-1-type) X-ray binaries, in which the optical star fills its Roche lobe, these systems were called Be/XR binaries.

The optical radiation of such a system is mainly produced by the Be star and the surrounding envelope. An accretion disk (AD) around the neutron star may also be a contributor. Gnedin *et al.* [4] were the first to have attempted to separate the additional (relative to the Be star) optical and infrared (*UBVRIJHK*) radiation and concluded that this radiation could be interpreted as the radiation of an AD whose parameters vary with time. Lyuty *et al.* [5] noticed that the object's X-ray activity was related to the global outburst of the Be star which began in late 1960s–early 1970s and reached a maximum in 1975, when the first of the three giant X-ray outbursts was observed. Such an outburst is attributable to a variable rate of mass loss from the Be star.

The luminous blue stars are known (see, e.g., [6]) to be variable in wide time and amplitude ranges. A moderate variability of $\Delta V \simeq 1^m$ is observed in most of these stars and is caused by the ejection of a dense envelope, which can be assumed to be the stellar pseudophotosphere. In this case, the bolometric luminosity is constant, but the pseudophotospheric temperature decreases and the radius increases. Since the star rapidly rotates (hundreds of km s$^{-1}$ on the equator), a disklike, rather than spherically symmetric envelope, is formed.

* e-mail: lyuty@sai.crimea.ua

Clark *et al.* [1, 7] investigated the spectroscopic and photometric variability of V725 Tau by using the 1983–1997 observations and showed that the optical variability was apparently attributable to a disklike envelope around the OB star (circumstellar disk), which is optically thin in *UBV*. The light variations are caused by variations of the emission measure in the circumstellar disk that emits by the free-free and bound-free mechanisms. There is no evidence for the existence of a second variable source (AD). A transient AD appears to be formed around the neutron star during giant outbursts [8]. In particular, the existence of an AD during the giant outburst of 1994 was proven [9]. However, the contribution of the AD to the optical luminosity of V725 Tau is very small and apparently does not exceed a few percent.

In this paper, we present our *UBV* observations of the star at the Crimean Station of the Sternberg Astronomical Institute (SAI) in 1980–1998 and analyze its photometric behavior over a period of 100 years and the color characteristics of the additional (relative to the Be star) radiation responsible for the optical variability.

## OBSERVATIONS

Our observations began in the spring of 1983 and were carried out with a photon-counting *UBV* photometer attached to the 60-m telescope of the Crimean Station of the SAI (pos. Nauchnyi). The instrumental photometric system was very close to Johnson's standard photometric system, as suggested by the mean transformation coefficients that were determined during six observing seasons in 1967–1990 from observations of standard stars in the Pleiades, Praesepe, and IC 4665 open clusters: $k_1 = -0.045 \pm 0.019$, $k_2 = 0.999 \pm 0.012$, $k_3 = 1.013 \pm 0.014$ [the corresponding formulas are $\Delta V = \Delta v + k_1 \Delta(B-V)$, $\Delta(B-V) = k_2 \Delta(b-v)$, and $\Delta(U-B) = k_3 \Delta(u-b)$]. As we see, $k_1$ is close to zero, while $k_2$ and $k_3$ are close to unity.

As the local standard, we used star "e" from [10], in which the following magnitudes are given for it: $V = 10^m\!.54 \pm 0^m\!.03$, $B-V = +0^m\!.45 \pm 0^m\!.03$, and $U-B = +0^m\!.19 \pm 0^m\!.05$. This comparison star was also used by Rössiger [11], who obtained its magnitudes similar to those of Lenouvel and Flogèr [10]: $V = 10^m\!.54$, $B-V = +0^m\!.45$, $U-B = +0^m\!.22$. Our measurements in 1984 and 1997 yielded the same magnitudes as those obtained by Lenouvel and Flogèr [10] but with a higher accuracy: $V = 10^m\!.53 \pm 0^m\!.01$, $B-V = +0^m\!.45 \pm 0^m\!.01$, and $U-B = +0^m\!.18 \pm 0^m\!.02$. As we see, the measurements of different authors over a period of 40 years yield the same magnitudes. Only the measurements by Walker in 1958 [11] differ markedly (up to $0^m\!.1$) from them; they are most likely erroneous.

Some of our observations have already been published [4, 5]; here, we present only the observations after 1988, as well as the observations that were performed by A.A. Aslanov in 1980 at the same site and with the same telescope and photometer and that he put at our disposal (see the table). Since Aslanov used the comparison star BD+26°876, we had to determine its magnitudes by referencing to the same standard (SA 48-3) which we used to determine the magnitudes of star "e" and reduce Aslanov's observations to our comparison star. Thus, Aslanov's observations constitute a single, homogeneous series with our measurements but are separated from ours by a period of three years and overlap in time with the observations by other authors. In our analysis, we therefore used Aslanov's observations together with published data.

## LIGHT AND COLOR CURVES

Portions of the light and color curves for the object have been published at different times. The most complete curves for a long time interval are given in [1, 5, 7]. We now have an opportunity to present the photoelectric light and color curves for A0535+26/V725 over a period of 25 years, from 1973 until 1998. The first systematic *UBV* observations of the object, which was not known at that time as an X-ray source, were initiated by Rössiger [11]. In their previous study, Rössiger and Wenzel [12], who used this star as one of the comparison stars for the observations of RR Tau, noticed its weak variability. These observations, together with the observations by De Loor *et al.* [13] and Aslanov (see the table), are the main photometric *UBV* data for V725 Tau before 1983. We did not use the observations of other authors after 1983, because, first, they could violate the homogeneity of the series (see [5]) and, second, they are rather few in number compared to our observations, while our observations represent satisfactorily the star's variability in 1983–1998.

Since here we are primarily interested in the global variability of the object, while the star also exhibits a rapid variability with a significant amplitude, up to $0^m\!.1$ in two or three days and up to $0^m\!.05$ during the night, we averaged the measurements close in time and/or brightness. These average magnitudes and colors are shown in Fig. 1, where the arrows mark the times of giant X-ray outbursts (we have in mind only the outbursts observed in the intermediate energy band 2–10 keV). Occasionally, the outburst in April 1989 is regarded as a giant outburst [2], but its spectrum was very hard: from 0.6 Crab in the band 2–6 keV to 4.3 Crab in the band 16–26 keV [14]. In the intermediate energy band 2–10 keV, it should apparently be considered "normal." This outburst is marked in Fig. 1 by the dashed arrow.

Several features can be noted in the behavior of the star's light and color during 25 years (see Fig. 1):

(1) On the average, the brightness declined from $8\overset{m}{.}8$–$8\overset{m}{.}9$ to $9\overset{m}{.}45$; the $B$–$V$ color index decreased, while $U$–$B$ first decreased but after 1995 began to increase.

(2) Three brightness states can be distinguished: (i) the high brightness in 1973–1981, the colors do not correlate with brightness; (ii) the pulsations in 1983–1995 at a nearly constant mean brightness, $B$–$V$ varies in antiphase with phase with brightness, and $U$–$B$ does not correlate with brightness; and (iii) the brightness decline after 1995 with a small outburst in 1996; $B$–$V$ also anticorrelates with brightness, while $U$–$B$ anticorrelates with $B$–$V$.

(3) The transitions between these states roughly coincide in time with giant outbursts; before the 1975 outburst, the $V$ brightness was constant, while the colors varied appreciably; after the 1980 outburst, the brightness declined by more than $0\overset{m}{.}2$, but the color indices were nearly constant; after the 1994 outburst, the colors varied in antiphase—the star became bluer in $B$–$V$ but redder in $U$–$B$.

In the fall of 1997, the star's brightness had declined to the level of the 1950s for the first time since 1973 and has remained approximately at this level to the present day (late 1998), undergoing only short-term fluctuations with an amplitude up to $0\overset{m}{.}1$. The dereddened colors correspond to a B0 star (see below for details on the colors). The lowest brightness was recorded in 1998, at which the colors $B$–$V$ = +0.425 and $U$–$B$ = –0.604 turned out to be similar to the colors in the mid-1950s, before the outburst [15, 10]: $B$–$V$ = +0.45 and +0.46, $U$–$B$ = –0.54 and –0.53, respectively.

PHOTOMETRIC HISTORY OF THE STAR

For the subsequent analysis, it is of interest to trace the photometric history of the star. Stier and Liller [16] gave yearly means of the Harvard photographic observations since 1898 and noted that the fading in 1940–1955 may have been a real effect. Rössiger [17] gave the Sonneberg photographic observations of the star since 1930. The Harvard and Sonneberg observations overlap in time, but their means differ markedly in the same years. For the calibration, we used the photoelectric observations by Hiltner [15] and Lenouvel and Flogèr [10] by assuming $B_{pg} = m_{pg} - 0.05$ for the Sonneberg observations and $B_{pg} = m_{pg} + 0.18$ for the Harvard observations before 1970 (the Harvard observations in 1973 and 1974 have a zero correction, which may be explained by the already commenced outburst of the star).

The archived photographic observations reduced in this way, together with the seasonally averaged photoelectric data, are shown in Fig. 2 and give an idea of the star's photometric behavior over a period of 100 years. The secular brightness decline from 1898 until 1950 is particularly noteworthy. The minimum in 1945–1949 is real—it is confirmed by the Sonneberg data. It is unlikely that the rapid (of the order of a year) fluctuations are real: first, the amplitude of these fluctuations is within the accuracy of photographic photometry, and, second, the Harvard and Sonneberg data disagree. In particular, the "outburst" of 1933 as inferred from the Harvard data is in conflict with the Sonneberg data.

In 1950–1965, the star was in quiescence; its brightness was at the mean level of $B = 9\overset{m}{.}83$ (indicated by the dashed line in Fig. 2), and then a fairly rapid brightness rise to the maximum of 1975–1980 began. We clearly see in Fig. 2 that the giant outbursts are not accidental but are associated with certain features of the light curve. The first outburst in 1975 occurred after some brightness level near the maximum was reached (maximum in $V$); the second outburst (1980) occurred before an abrupt brightness decline, after which quasi-periodic (~1400 days) fluctuations [7] set in at a nearly constant mean level; and after the third outburst (1994), the amplitude of these fluctuations abruptly decreased (the maximum of 1996 agrees with a quasi-period of 1400 days), and the brightness declined to the 1950–1965 level in 1997–1998 and below this level in the fall of 1998. The filled square in Fig. 2 marks *minimum minimorum*, the lowest brightness for the entire period of photoelectric observations, which was observed in the fall of 1998. It may well be that the same minimum as that in 1945–1949 will be observed in the next several years.

Thus, at least two different states of the star were observed during the century: quiescence (1898–1965) and an outburst—an active phase (1970–1997) with the amplitude $\Delta V = 0\overset{m}{.}55$. This behavior is characteristic of Be stars with a moderate variability and is explained by the ejection of an envelope which produces a pseudophotosphere [6]. In this case, the outburst amplitude is determined by a change in the mass-loss rate $\Delta \log \dot{M} \approx -0.3\Delta V$. The bolometric luminosity is nearly constant, the pseudophotospheric temperature decreases compared to the photosphere of the Be star, and its radius increases.

ADDITIONAL RADIATION

*Mean Color Indices*

To elucidate the nature of the optical variability in A0535+26/V725 Tau, it is first necessary to separate the additional (relative to the Be star) radiation and determine its characteristics. For this purpose, the intrinsic $UBV$ magnitudes of the star must be known. We could take the 1953–1956 magnitudes as the main brightness, but the star's mean brightness in the fall of 1997 declined below this level. In the fall of 1998, we managed to record the minimum brightness in all

Photoelectric observations of A0536+26/V725 Tau in 1980 and 1988–1998

| JD 2400000+ | V | B–V | U–B | JD 2400000+ | V | B–V | U–B |
|---|---|---|---|---|---|---|---|
| 44254.526 | 8.87 | 0.54 | −0.60 | 48158.519 | 9.311 | 0.415 | −0.670 |
| 44257.334 | 8.86 | 0.56 | −0.65 | 48158.562 | 9.307 | 0.421 | −0.678 |
| 44261.217 | 8.864 | 0.536 | −0.568 | 48164.570 | 9.320 | 0.430 | −0.655 |
| 44488.560 | 8.878 | 0.562 | −0.61 | 48176.523 | 9.217 | 0.460 | −0.656 |
| 44489.555 | 8.868 | 0.552 | −0.61 | 48188.519 | 9.286 | 0.433 | −0.694 |
| 44490.563 | 8.863 | 0.567 | −0.63 | 48189.452 | 9.286 | 0.447 | −0.690 |
| 44493.557 | 8.86 | 0.56 | −0.55 | 48211.404 | 9.218 | 0.464 | −0.696 |
| 44494.576 | 8.88 | 0.57 | −0.62 | 48211.417 | 9.218 | 0.468 | −0.684 |
| 44589.515 | 8.88 | 0.57 | −0.625 | 48244.268 | 9.295 | 0.447 | −0.677 |
| 44590.401 | 8.89 | 0.56 | −0.69 | 48251.265 | 9.308 | 0.424 | −0.678 |
| 44599.401 | 8.93 | 0.56 | −0.72 | 48251.277 | 9.305 | 0.431 | −0.672 |
| 44600.469 | 8.88 | 0.60 | −0.72 | 48252.233 | 9.275 | 0.431 | −0.718 |
| 47417.559 | 9.140 | 0.475 | −0.645 | 48272.206 | 9.321 | 0.424 | −0.658 |
| 47418.570 | 9.110 | 0.491 | −0.666 | 48274.222 | 9.317 | 0.428 | −0.678 |
| 47419.554 | 9.090 | 0.509 | −0.666 | 48294.240 | 9.251 | 0.451 | −0.704 |
| 47420.577 | 9.092 | 0.497 | −0.670 | 48295.291 | 9.257 | 0.452 | −0.673 |
| 47439.578 | 9.127 | 0.499 | −0.669 | 48325.251 | 9.157 | 0.477 | −0.672 |
| 47449.542 | 9.093 | 0.502 | −0.638 | 48505.565 | 9.107 | 0.513 | −0.689 |
| 47450.574 | 9.091 | 0.492 | −0.667 | 48508.573 | 9.112 | 0.496 | −0.704 |
| 47451.554 | 9.077 | 0.489 | −0.625 | 48509.575 | 9.114 | 0.503 | −0.704 |
| 47536.342 | 9.001 | 0.516 | −0.616 | 48515.568 | 9.088 | 0.516 | −0.701 |
| 47540.322 | 9.005 | 0.515 | −0.626 | 48516.573 | 9.084 | 0.519 | −0.698 |
| 47569.266 | 8.994 | 0.515 | −0.634 | 48525.585 | 9.056 | 0.525 | −0.693 |
| 47586.256 | 8.982 | 0.523 | −0.628 | 48534.592 | 9.104 | 0.501 | −0.680 |
| 47607.248 | 8.974 | 0.531 | −0.620 | 48539.565 | 9.118 | 0.524 | −0.689 |
| 47624.263 | 9.014 | 0.543 | −0.629 | 48540.472 | 9.122 | 0.501 | −0.698 |
| 47834.615 | 9.231 | 0.459 | −0.714 | 48541.557 | 9.112 | 0.511 | −0.701 |
| 47835.485 | 9.240 | 0.437 | −0.708 | 48543.512 | 9.108 | 0.503 | −0.714 |
| 47836.445 | 9.219 | 0.478 | −0.700 | 48546.451 | 9.131 | 0.508 | −0.719 |
| 47837.557 | 9.251 | 0.462 | −0.698 | 48566.406 | 9.136 | 0.517 | −0.723 |
| 47838.472 | 9.286 | 0.451 | −0.716 | 48576.472 | 9.155 | 0.506 | −0.718 |
| 47855.450 | 9.227 | 0.441 | −0.699 | 48603.369 | 9.218 | 0.472 | −0.722 |
| 47861.394 | 9.238 | 0.459 | −0.702 | 48856.520 | 9.005 | 0.511 | −0.616 |
| 47943.228 | 9.098 | 0.491 | −0.651 | 48861.526 | 8.988 | 0.513 | −0.599 |
| 47943.235 | 9.102 | 0.491 | −0.653 | 48864.526 | 8.975 | 0.524 | −0.644 |
| 47944.225 | 9.115 | 0.486 | −0.664 | 48865.496 | 8.993 | 0.513 | −0.615 |
| 47944.238 | 9.113 | 0.486 | −0.656 | 48866.498 | 8.991 | 0.536 | −0.622 |
| 47946.224 | 9.132 | 0.491 | −0.686 | 48870.500 | 9.011 | 0.510 | −0.614 |
| 47946.230 | 9.125 | 0.490 | −0.690 | 48893.502 | 8.989 | 0.528 | −0.636 |
| 47948.233 | 9.134 | 0.487 | −0.683 | 48894.598 | 8.989 | 0.509 | −0.645 |
| 47966.260 | 9.162 | 0.486 | −0.696 | 48916.497 | 8.981 | 0.529 | −0.653 |
| 47968.269 | 9.150 | 0.484 | −0.668 | 48930.521 | 9.008 | 0.526 | −0.642 |
| 47970.258 | 9.172 | 0.466 | −0.656 | 48958.345 | 9.004 | 0.527 | −0.658 |
| 47971.276 | 9.178 | 0.489 | −0.721 | 48959.335 | 8.993 | 0.528 | −0.651 |
| 47981.243 | 9.157 | 0.487 | −0.702 | 48961.315 | 8.988 | 0.538 | −0.661 |
| 47984.291 | 9.139 | 0.461 | −0.656 | 48992.222 | 9.096 | 0.508 | −0.717 |
| 47986.286 | 9.155 | 0.479 | −0.713 | 49002.212 | 9.090 | 0.523 | −0.687 |
| 48151.565 | 9.348 | 0.421 | −0.650 | 49006.254 | 9.093 | 0.517 | −0.727 |

**Table.** (Contd.)

| JD 2400000+ | V | B–V | U–B | JD 2400000+ | V | B–V | U–B |
|---|---|---|---|---|---|---|---|
| 49031.317 | 9.092 | 0.487 | −0.687 | 49973.552 | 9.333 | 0.435 | −0.676 |
| 49059.275 | 9.092 | 0.507 | −0.691 | 49973.559 | 9.343 | 0.424 | −0.678 |
| 49062.282 | 9.108 | 0.517 | −0.681 | 49998.462 | 9.339 | 0.420 | −0.731 |
| 49245.553 | 9.024 | 0.524 | −0.685 | 49998.468 | 9.342 | 0.426 | −0.727 |
| 49246.561 | 9.033 | 0.532 | −0.680 | 50002.435 | 9.279 | 0.442 | −0.628 |
| 49248.540 | 9.028 | 0.534 | −0.672 | 50002.446 | 9.273 | 0.447 | −0.649 |
| 49251.571 | 9.038 | 0.535 | −0.684 | 50002.452 | 9.278 | 0.440 | −0.654 |
| 49253.557 | 9.049 | 0.528 | −0.672 | 50004.424 | 9.254 | 0.457 | −0.661 |
| 49255.568 | 9.041 | 0.524 | −0.682 | 50006.436 | 9.283 | 0.440 | −0.676 |
| 49273.544 | 9.029 | 0.528 | −0.687 | 50007.423 | 9.294 | 0.444 | −0.673 |
| 49274.597 | 9.017 | 0.518 | −0.680 | 50007.434 | 9.294 | 0.443 | −0.669 |
| 49359.261 | 9.187 | 0.476 | −0.689 | 50008.418 | 9.309 | 0.441 | −0.663 |
| 49360.323 | 9.193 | 0.475 | −0.683 | 50008.430 | 9.312 | 0.437 | −0.669 |
| 49373.411 | 9.198 | 0.474 | −0.683 | 50009.392 | 9.308 | 0.430 | −0.666 |
| 49381.228 | 9.224 | 0.459 | −0.712 | 50009.404 | 9.310 | 0.439 | −0.675 |
| 49393.293 | 9.241 | 0.458 | −0.677 | 50010.431 | 9.332 | 0.429 | −0.667 |
| 49394.297 | 9.244 | 0.455 | −0.687 | 50010.444 | 9.326 | 0.436 | −0.675 |
| 49400.322 | 9.254 | 0.441 | −0.667 | 50014.394 | 9.334 | 0.442 | −0.677 |
| 49401.290 | 9.249 | 0.454 | −0.680 | 50014.406 | 9.332 | 0.435 | −0.658 |
| 49407.217 | 9.274 | 0.472 | −0.696 | 50015.428 | 9.348 | 0.439 | −0.674 |
| 49412.260 | 9.273 | 0.442 | −0.670 | 50015.442 | 9.359 | 0.428 | −0.669 |
| 49438.231 | 9.182 | 0.480 | −0.676 | 50047.385 | 9.326 | 0.434 | −0.678 |
| 49442.244 | 9.158 | 0.486 | −0.673 | 50047.396 | 9.313 | 0.449 | −0.688 |
| 49444.254 | 9.181 | 0.475 | −0.656 | 50048.408 | 9.307 | 0.448 | −0.678 |
| 49450.252 | 9.147 | 0.484 | −0.652 | 50048.442 | 9.293 | 0.439 | −0.688 |
| 49457.244 | 9.183 | 0.490 | −0.650 | 50048.455 | 9.288 | 0.443 | −0.682 |
| 49596.538 | 9.300 | 0.450 | −0.695 | 50064.386 | 9.293 | 0.436 | −0.702 |
| 49600.543 | 9.324 | 0.462 | −0.685 | 50064.418 | 9.286 | 0.439 | −0.681 |
| 49601.518 | 9.335 | 0.434 | −0.687 | 50064.431 | 9.288 | 0.442 | −0.687 |
| 49609.556 | 9.344 | 0.433 | −0.648 | 50093.468 | 9.233 | 0.463 | −0.721 |
| 49611.562 | 9.336 | 0.442 | −0.665 | 50093.475 | 9.195 | 0.465 | −0.678 |
| 49629.460 | 9.292 | 0.431 | −0.646 | 50095.488 | 9.206 | 0.469 | −0.675 |
| 49630.552 | 9.295 | 0.435 | −0.654 | 50102.287 | 9.255 | 0.454 | −0.693 |
| 49640.540 | 9.242 | 0.457 | −0.682 | 50106.298 | 9.249 | 0.452 | −0.688 |
| 49653.418 | 9.297 | 0.446 | −0.658 | 50108.194 | 9.278 | 0.458 | −0.691 |
| 49654.470 | 9.300 | 0.450 | −0.680 | 50115.299 | 9.267 | 0.437 | −0.746 |
| 49658.379 | 9.320 | 0.448 | −0.664 | 50116.369 | 9.265 | 0.435 | −0.680 |
| 49663.462 | 9.316 | 0.440 | −0.665 | 50117.193 | 9.271 | 0.441 | −0.695 |
| 49665.460 | 9.335 | 0.420 | −0.686 | 50121.303 | 9.236 | 0.454 | −0.692 |
| 49692.367 | 9.296 | 0.426 | −0.664 | 50133.343 | 9.250 | 0.465 | −0.688 |
| 49716.333 | 9.302 | 0.435 | −0.670 | 50139.215 | 9.241 | 0.472 | −0.700 |
| 49717.328 | 9.314 | 0.434 | −0.664 | 50139.240 | 9.251 | 0.459 | −0.734 |
| 49737.238 | 9.315 | 0.445 | −0.674 | 50155.221 | 9.273 | 0.476 | −0.677 |
| 49751.351 | 9.313 | 0.461 | −0.669 | 50170.298 | 9.28 | 0.50 | −0.71 |
| 49771.230 | 9.353 | 0.420 | −0.648 | 50177.258 | 9.231 | 0.449 | −0.770 |
| 49772.232 | 9.364 | 0.417 | −0.647 | 50178.256 | 9.259 | 0.439 | −0.705 |
| 49778.244 | 9.361 | 0.422 | −0.663 | 50311.535 | 9.157 | 0.508 | −0.682 |
| 49782.232 | 9.382 | 0.414 | −0.641 | 50361.468 | 9.119 | 0.497 | −0.703 |
| 49820.270 | 9.306 | 0.446 | −0.666 | 50362.516 | 9.161 | 0.500 | −0.701 |
| 49820.278 | 9.311 | 0.442 | −0.670 | 50373.542 | 9.157 | 0.497 | −0.719 |
| 49825.262 | 9.286 | 0.470 | −0.667 | 50373.558 | 9.152 | 0.504 | −0.724 |

**Table.** (Contd.)

| JD 2400000+ | V | B–V | U–B | JD 2400000+ | V | B–V | U–B |
|---|---|---|---|---|---|---|---|
| 50392.604 | 9.230 | 0.469 | −0.711 | 50760.384 | 9.451 | 0.412 | −0.618 |
| 50393.595 | 9.226 | 0.477 | −0.711 | 50760.415 | 9.462 | 0.398 | −0.602 |
| 50393.607 | 9.232 | 0.485 | −0.714 | 50761.426 | 9.448 | 0.406 | −0.602 |
| 50395.513 | 9.224 | 0.480 | −0.721 | 50762.423 | 9.443 | 0.405 | −0.617 |
| 50395.537 | 9.229 | 0.476 | −0.714 | 50762.458 | 9.436 | 0.407 | −0.596 |
| 50397.469 | 9.227 | 0.461 | −0.715 | 50783.483 | 9.445 | 0.403 | −0.626 |
| 50400.465 | 9.240 | 0.473 | −0.707 | 50793.310 | 9.455 | 0.413 | −0.618 |
| 50400.498 | 9.229 | 0.470 | −0.705 | 50802.348 | 9.418 | 0.407 | −0.642 |
| 50402.553 | 9.228 | 0.483 | −0.734 | 50817.248 | 9.472 | 0.396 | −0.604 |
| 50403.452 | 9.244 | 0.463 | −0.708 | 50818.227 | 9.469 | 0.398 | −0.595 |
| 50405.392 | 9.199 | 0.479 | −0.716 | 50827.296 | 9.477 | 0.409 | −0.573 |
| 50408.510 | 9.257 | 0.455 | −0.705 | 50863.287 | 9.442 | 0.402 | −0.610 |
| 50478.279 | 9.295 | 0.434 | −0.676 | 50867.304 | 9.465 | 0.413 | −0.629 |
| 50484.356 | 9.304 | 0.442 | −0.695 | 50868.246 | 9.470 | 0.404 | −0.617 |
| 50487.282 | 9.344 | 0.430 | −0.685 | 50875.291 | 9.452 | 0.407 | −0.600 |
| 50491.308 | 9.350 | 0.434 | −0.715 | 50920.246 | 9.449 | 0.417 | −0.617 |
| 50491.320 | 9.356 | 0.427 | −0.717 | 50920.252 | 9.456 | 0.420 | −0.641 |
| 50504.347 | 9.357 | 0.424 | −0.689 | 51044.551 | 9.461 | 0.419 | −0.585 |
| 50505.324 | 9.370 | 0.425 | −0.674 | 51046.548 | 9.472 | 0.445 | −0.628 |
| 50509.303 | 9.354 | 0.430 | −0.669 | 51052.553 | 9.399 | 0.442 | −0.603 |
| 50510.324 | 9.350 | 0.429 | −0.646 | 51054.551 | 9.475 | 0.417 | −0.619 |
| 50510.338 | 9.357 | 0.425 | −0.646 | 51061.544 | 9.458 | 0.421 | −0.604 |
| 50519.296 | 9.376 | 0.419 | −0.664 | 51074.569 | 9.448 | 0.414 | −0.603 |
| 50519.308 | 9.369 | 0.423 | −0.649 | 51074.582 | 9.460 | 0.405 | −0.606 |
| 50521.322 | 9.377 | 0.425 | −0.655 | 51075.513 | 9.430 | 0.422 | −0.605 |
| 50526.313 | 9.363 | 0.426 | −0.687 | 51075.521 | 9.437 | 0.414 | −0.600 |
| 50689.558 | 9.390 | 0.416 | −0.624 | 51081.504 | 9.441 | 0.417 | −0.602 |
| 50689.565 | 9.387 | 0.419 | −0.636 | 51081.517 | 9.461 | 0.395 | −0.599 |
| 50690.552 | 9.344 | 0.425 | −0.628 | 51088.536 | 9.477 | 0.410 | −0.626 |
| 50696.562 | 9.414 | 0.406 | −0.614 | 51103.549 | 9.452 | 0.400 | −0.618 |
| 50699.566 | 9.404 | 0.412 | −0.621 | 51103.555 | 9.492 | 0.388 | −0.628 |
| 50704.560 | 9.425 | 0.412 | −0.612 | 51104.479 | 9.473 | 0.403 | −0.614 |
| 50726.502 | 9.421 | 0.390 | −0.617 | 51104.492 | 9.472 | 0.403 | −0.605 |
| 50730.460 | 9.445 | 0.405 | −0.609 | 51105.405 | 9.472 | 0.404 | −0.589 |
| 50748.452 | 9.441 | 0.405 | −0.619 | 51110.382 | 9.475 | 0.407 | −0.600 |
| 50748.463 | 9.450 | 0.398 | −0.611 | 51111.443 | 9.475 | 0.408 | −0.622 |
| 50751.592 | 9.444 | 0.398 | −0.602 | 51112.482 | 9.468 | 0.401 | −0.615 |
| 50753.459 | 9.430 | 0.401 | −0.606 | 51112.493 | 9.464 | 0.412 | −0.620 |
| 50753.470 | 9.444 | 0.394 | −0.598 | 51137.350 | 9.408 | 0.418 | −0.623 |
| 50754.586 | 9.446 | 0.402 | −0.599 | 51141.364 | 9.448 | 0.394 | −0.611 |
| 50754.618 | 9.440 | 0.408 | −0.603 | 51141.369 | 9.438 | 0.415 | −0.632 |
| 50755.447 | 9.440 | 0.397 | −0.623 | 51163.284 | 9.418 | 0.412 | −0.607 |
| 50755.474 | 9.440 | 0.408 | −0.623 | 51163.288 | 9.413 | 0.410 | −0.613 |
| 50758.430 | 9.435 | 0.413 | −0.620 | 51164.382 | 9.448 | 0.414 | −0.610 |
| 50758.455 | 9.431 | 0.415 | −0.604 | 51164.386 | 9.443 | 0.418 | −0.624 |
| 50759.428 | 9.449 | 0.405 | −0.612 | 51175.276 | 9.466 | 0.408 | −0.603 |
| 50759.454 | 9.446 | 0.401 | −0.598 | 51175.281 | 9.472 | 0.408 | −0.611 |

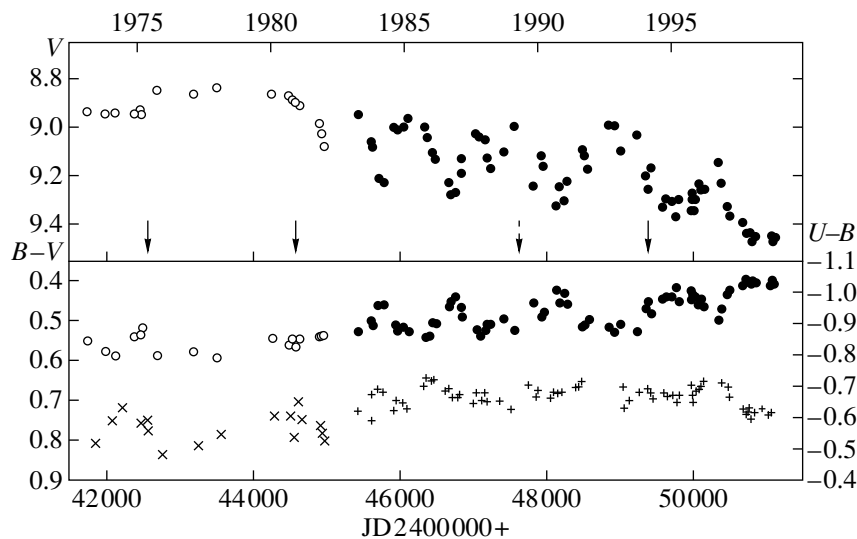

**Fig. 1.** Smoothed light and color curves of V725 Tau/A0535+26: the filled circles and pluses are our observations in 1983–1998, and the open circles and crosses are the observations of other authors before 1983 ([11, 35, 13]; Aslanov, see the table). The arrows mark the times of giant X-ray outbursts.

26 years of photoelectric observations (*minimum minimorum*). Apparently because of the rapid variability in the star, the minimum brightness in different bands was observed at different times and was found to be $V_{min} = 9^{m}\!.492 \pm 0^{m}\!.005$, $B_{min} = 9^{m}\!.917 \pm 0^{m}\!.005$, and $U_{min} = 9^{m}\!.313 \pm 0^{m}\!.008$; the colors at this minimum were $B-V = 0^{m}\!.425$ and $U-B = -0^{m}\!.604$.

Next, the colors must be corrected for the reddening, which is rather large for V725 Tau. Margon *et al.* [18] and Wade and Oke [19] gave the color excesses $E_{B-V} = 0^{m}\!.8$ and $0^{m}\!.80$, respectively, while Giangrande *et al.* [3] obtained $E_{B-V} = 0^{m}\!.82 \pm 0^{m}\!.04$ and $E_{U-B} = 0^{m}\!.55 \pm 0^{m}\!.04$. However, these values appear to be slightly overestimated. Since the luminosity class of the star was found to be III–V, the dereddened colors must correspond to the main sequence. If the radiation from the B star proper is assumed to have been observed in 1953–1956 and in the fall of 1998 (*minimum minimorum*), then we obtain the reddenings $E_{B-V} = 0^{m}\!.74$ and $E_{U-B} = 0^{m}\!.48$. In this case, the color excess $E_{B-V}$ matches the value $0^{m}\!.75 \pm 0^{m}\!.05$ obtained from the 2200 Å interstellar band [2], while $E_{U-B}$ is slightly smaller than that followed from the normal reddening law. Below, we use our color excesses.

Thus, using the *minimum minimorum* magnitudes and our color excesses, we can derive the intrinsic colors of the additional radiation, which determined the outburst (active phase) of the star in 1973–1997 (see Fig. 2). In Fig. 3, the colors of the additional radiation are plotted against observed brightness. The $(B-V)_{add}$ color index monotonically decreases with declining brightness, while $(U-B)_{add}$ first decreases and then begins to increase. Comparing Fig. 3 with the light curve in Fig. 1, we see that the rise in brightness by $\Delta V \simeq 0^{m}\!.1$ from 1973 until 1977 and its subsequent decline after 1980 had no effect on the dependence in Fig. 3: The colors of the additional radiation depend only on brightness and not on the time when this brightness was reached. Only one data point deviates appreciably from the mean dependence of $(U-B)_{add}$ on $V_{obs}$ (it corresponds to the minimum in late 1981–early 1982; see Figs. 1 and 2). It thus follows that all the additional radiation of 1973–1997 was attributable to a single source, apparently the circumstellar envelope (recall once again that here we do not consider rapid fluctuations of a relatively small amplitude, which may have a different nature).

Noteworthy is the large scatter for $(U-B)_{add}$, which is much larger than the observational errors and is mostly likely attributable to the Balmer emission jump.

### Variability of the Hα Emission Line

If the additional radiation originates in the circumstellar envelope, then the intensity of emission lines, primarily Hα, must correlate with brightness. Such a correlation was actually found [1]: The Hα equivalent width as deduced from the 1987–1996 data correlates with V magnitude and B–V color at a 95% confidence level. It should be noted, however, that the continuum also varies; i.e., the equivalent width of the line reflects not only the variability of its intensity but also the continuum variability. Allowance for the continuum variability must therefore increase the confidence level of the correlation. In particular, if the continuum variability is taken into account, the correlation coefficient for the 1987–1996 data increases by more than a factor of 1.5, and the confidence level increases to 99.9%.

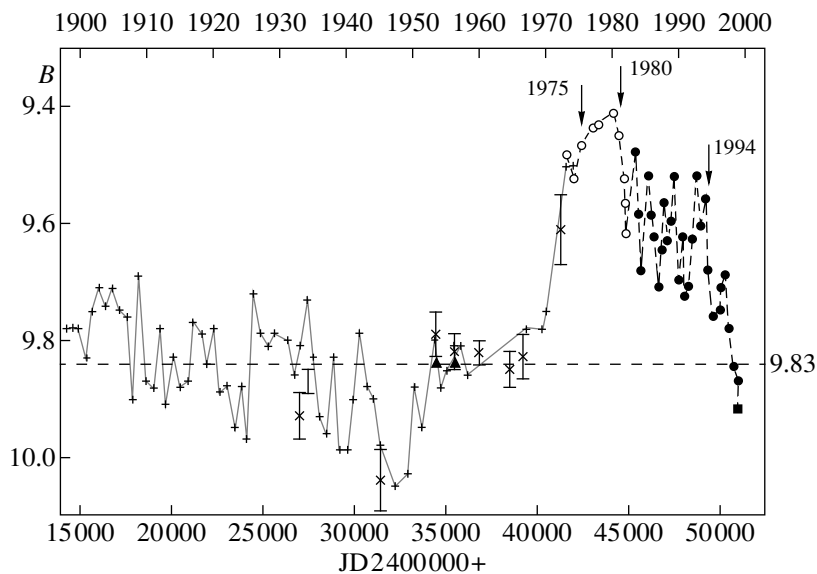

**Fig. 2.** Photometric history of the star from 1898: The pluses are the Harvard photographic observations [16], the crosses are the Sonneberg observations [17], the triangles are the photoelectric observations by Hiltner [15] and Lenouvel and Flogèr [10], and the square is the *minimum minimorum* of 1998; the remaining notation is the same as that in Fig. 1.

The maximum brightness was observed in 1975–1980 (Fig. 2), and the first spectroscopic observations began approximately at the same time. We collected all published data on Hα equivalent-width determinations from 1975 until 1981 [13, 19–24] and reduced all values to the same continuum $V = 9.^{m}50$ by assuming that the continuum variations near Hα were proportional to those in $V$. This reduced equivalent width $EW_{red}$ is proportional to the line intensity. When the dates of observation in $V$ and in Hα did not coincide, we took the nearest date (but no farther than ten days) of $V$ measurements with allowance for the variability pattern (Fig. 1). These results, together with the data of Clark *et al.* [1], which were also reduced to $V = 9.^{m}50$, are shown in the upper panel of Fig. 3. The open circles represent the 1975–1981 data, and the filled circles represent the 1987–1996 and 1997–1998 data [25].

Figure 3 shows not only a clear dependence of the Hα intensity on brightness but also a different behavior of this dependence at different times. In 1975–1981 (open circles) the Hα intensity was, on the average, triple that in 1987–1996. There is no significant correlation between line intensity and brightness. However, Aab [24] noted a considerable increase in $EW(Hα)$ in January 1981, three months after the giant outburst of 1980, but it is hard to tell whether this increase is attributable to the outburst, because there were no spectroscopic observations during the outburst.

All 1987–1998 data show a good correlation with brightness; the correlation coefficient is $r = 0.70$. The Hα line in emission rather than in absorption corresponds to the lowest brightness in the fall of 1998 [25]. However, two different levels can be clearly identified in the 1987–1997 dependence: $EW_{red}(Hα) = -14.6 \pm 0.9$ for $V = 9.^{m}0–9.^{m}15$, $\bar{V} = 9.^{m}11$ and $EW_{red}(Hα) = -9.1 \pm 0.4$ for $V = 9.^{m}2–9.^{m}4$, $\bar{V} = 9.^{m}30$. The separation into these two levels depends not on time but only on brightness, which is apparently attributable to changes of the physical conditions in the envelope.

### Nature of the Additional Radiation

We now try to determine the characteristics and nature of this additional radiation, at least qualitatively. Since, as was already pointed out above, we are concerned with the global variations, let us smooth the brightness dependences of the colors of the additional radiation (Fig. 3) by a moving average with the interval $\Delta V = 0.^{m}1$ and with a shift by half the interval (a slightly smaller shift of the interval was used for a brightness lower than $9.^{m}3$). The positions of these mean colors of the additional radiation responsible for the optical variability in the two-color $(U–B)–(B–V)$ diagram are shown in Fig. 4, which also shows the radiation of a different nature: main-sequence stars, a blackbody, synchrotron radiation, and the radiation of an optically thin and optically thick (in the Balmer continuum) hydrogen plasma [26]. In order not to overload the figure, we give only the errors in $U–B$, because the errors in $B–V$ are approximately a factor of 2 smaller; the numbers denote different apparent magnitudes. In addition, Fig. 4 shows the star's positions as determined from the 1953–1956 observations before the outburst (triangles) and at *minimum minimorum* of 1998 (square). We clearly see that the additional radiation at maximum light $V_{obs} = 8.^{m}85$ corresponds best to an optically thick (in the Balmer continuum and in *UBV*) plasma with $T_e \sim 1.5 \times 10^4$ K (the colors are close to the blackbody line). It should be borne in mind, however, that the calculations were performed for a spherically symmetric envelope, while the envelope in our case is most likely disklike [7].

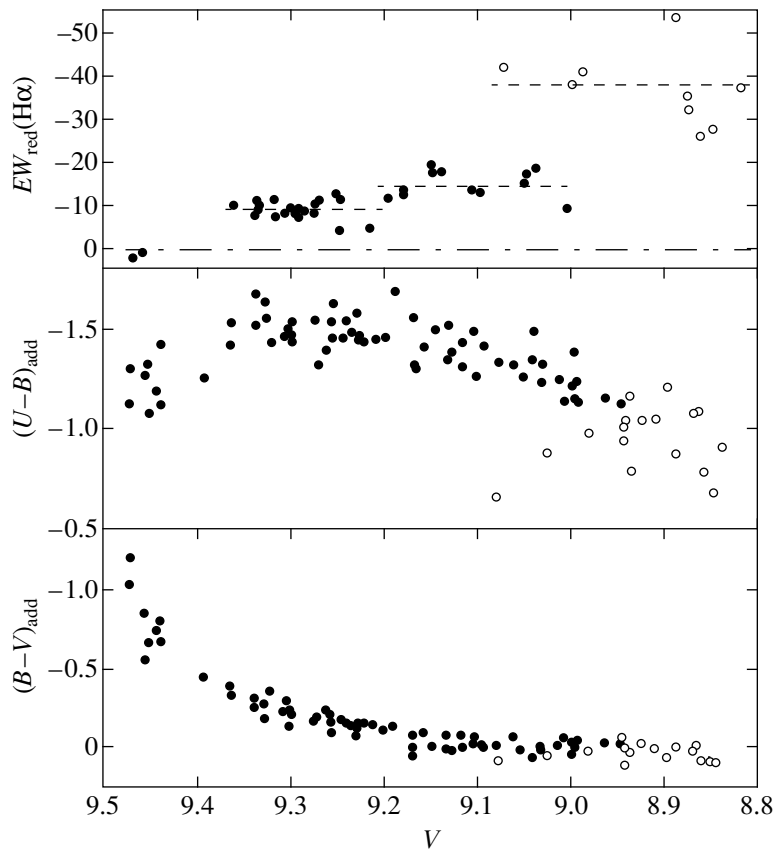

**Fig. 3.** Colors of the additional radiation in V725 Tau (the notation is the same as that in Fig. 1) and $EW(H\alpha)$ (in Angstroms) reduced to a single continuum level versus observed brightness. For $EW(H\alpha)$, the open and filled circles refer to 1975–1981 and 1987–1998, respectively.

The brightness decline corresponds to a decrease of the optical depth both in *UBV* and in the Balmer continuum; the colors approach the lines for an optically thin (in the Balmer continuum) plasma with $T_e = 10^4$ K and $N_e = 10^{10}$–$10^{12}$ cm$^{-3}$. The subsequent fading is apparently caused by a decrease of the optical depth in *UBV*. This behavior suggests the following natural scenario: An increase in the rate of mass loss from the Be star leads to the formation of a fairly dense envelope (pseudophotosphere), to a rise in brightness [6], and to the appearance of strong emission lines. In the course of time, the envelope begins to disperse, its optical depth decreases not only in the continuum but also in the lines, and the brightness and the emission-line intensity decrease. When the envelope becomes optically thin, it is no longer visible in *UBV* but apparently can still show up in the lines.

Indeed, at maximum light (1975–1980) the envelope was optically thick (see Figs. 2 and 4) and emitted radiation roughly as the stellar photosphere or, to be more precise, as a blackbody with $T_e \sim 12000$ K (pseudophotosphere) with a mean magnitude brighter than $V = 9.^{m}0$. The independence of $EW_{red}$ on brightness can be explained by line saturation. The physical conditions in the envelope appear to have changed abruptly after the giant outburst of 1980: Quasi-periodic fluctuations set in at a nearly constant mean brightness $V = 9.^{m}15$, the envelope became optically thin in the Balmer continuum, and, in general, a dependence of the H$\alpha$ intensity on brightness appeared. However, the two levels mentioned above, which may not accidentally correspond to the maximum and minimum of the 1400-day fluctuations, can be distinguished in this dependence. At a very low brightness (1998), the line is no longer in emission but in absorption: $EW_{red} \approx EW = 2.2$ Å (November 6, 1998). However, on November 9, 1998, a weak emission appeared on the red side, although, in general, the line remained in absorption. The line FWHM (~1000 km s$^{-1}$) was essentially the same as that for the emission [25].

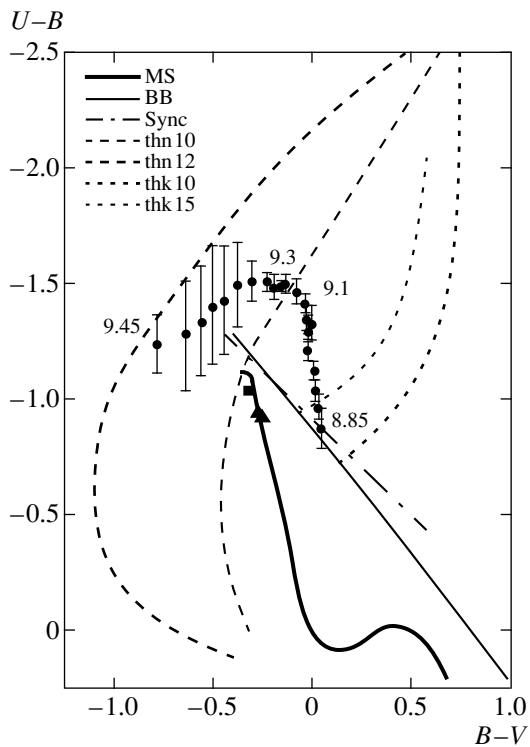

**Fig. 4.** Positions of the dereddened colors of the additional radiation in the $(U–B)$–$(B–V)$ diagram; the observed brightness is indicated for some levels. Also shown are the positions of main-sequence (MS) stars, a blackbody (BB), synchrotron radiation (Sync), an optically thin (in the Balmer continuum) hydrogen plasma with $T_e = 10^4$ K and $N_e = 10^{10}$ (thn10) and $N_e = 10^{12}$ cm$^{-3}$ (thn12), and an optically thick (in the Balmer continuum) plasma with $T_e = 10^4$ (thk10) and $T_e = 1.5 \times 10^4$ K (thk15). The filled triangles mark the star's positions in 1953–1956, and the filled square marks the *minimum minimorum* of 1998.

We now show that the additional radiation cannot be produced by the AD around the neutron star. An AD is known to emit radiation as a hot star with a temperature of $(2–3) \times 10^4$ K [27]. Furthermore, the outer disk regions can be additionally heated by X-ray radiation; the energy release through heating can exceed the gravitational one [28, 29]. Since the AD as a plane figure emits maximum radiation in the direction perpendicular to its plane, it gives the largest contribution to the flux from the binary system at small orbital inclinations. Such a situation is observed in the system ScoX-1 [29], where the disk contribution to the optical flux is at a maximum, but the radiation from the AD, the star, and a hot spot cannot be separated.

However, there is a binary system (HZ Her/Her X-1) where we can separate the AD radiation and determine its color characteristics. Both optical and X-ray eclipses are observed in HZ Her. The optical flux from this system has three components: (i) a normal A5–F star, (ii) a hot spot on the star's side facing the X-ray pulsar, and (iii) an AD around the neutron star. Since the neutron-star size is much smaller than the size of a normal A–F star, at the beginning and at the end of an X-ray eclipse, we see the radiation from the star itself and from half the AD but do not see the hot spot, which gives the bulk of the optical flux. During a total eclipse (phase 0.0), we see the flux only from the undisturbed side of the normal star. We can thus easily determine the AD colors observed at different eclipse phases [30].

In Fig. 5, the mean colors of the AD in HZ Her which correspond to different parts of the eclipsed disk region are indicated by the open diamonds; the upper diamond corresponds to half the disk at the beginning of the X-ray eclipse. The total colors of the AD are close to the colors of main-sequence B1–B5-stars (B5 corresponds to the outer, colder parts of the disk). The colors of the additional radiation in A0535+26/V725 Tau, first, are far from the AD colors and, second, exhibit a completely different pattern of variations with total flux. We therefore conclude that the contribution of the AD in this system is very small and unlikely exceeds a few percent.

This figure also shows the color variations of the additional radiation during the giant outbursts of 1980 and 1994 and the normal outburst of 1983. The X-ray outbursts are known to be divided into giant and normal outbursts only by the amplitude, but we do not know whether they are different in nature. However, Motch *et al.* [8] assume that a transient AD is formed around the neutron star during giant outbursts and that stellar-wind accretion takes place during normal outbursts. The first giant outburst occurred in late April–early May 1975, when the object's visibility season for ground-based observations had ended. The 1980 and 1994 outbursts were extensively observed in X rays and optically.

Numbers 1–5 without symbols in Fig. 5 indicate the color behavior of the additional radiation during the giant outburst of 1980 (G80). The color variations correspond to the variations both in temperature and in optical depth: A month before the outburst (1, 2), the temperature was higher, while the optical depth was smaller than those at the maximum (3); the minimum temperature and the maximum optical depth were observed ten days after the maximum (4); and a month after the maximum, the state was approximately the same (5) as a month before the maximum. The color variations during the normal outburst in October 1983 (N83, 6–9) were completely different in pattern. First, the circumstellar envelope became optically thin in the Balmer continuum, and, second, the optical depth in *UBV* decreased at the maximum (8). The colors during the giant outburst of 1994 (10–14) exhibited approximately the same behavior; i.e., the nature of the 1994 outburst is apparently the same as that of a normal outburst, but the nature of the outburst in 1980 (and most likely in 1975) is clearly completely different.

## DISCUSSION

This article is a continuation of the series of papers [1, 7, 31] on the study of long-term variability in the Be/XR binary system A0535+26/V725 Tau. Previously, the following conclusions were reached: The circumstellar envelope (disk) is responsible for the object's optical variability; there is no evidence for the presence of a second variable source (an AD around the neutron star); the long-term (years or, possibly, months) variability of $EW(H\alpha)$ correlates with the $V$ magnitude and the $B-V$ color, but there is no such correlation for $V/R$ (the ratio of the violet and red emission peaks), although the $V/R$ ratio varies in a fairly wide range, possibly periodically with a period of about a year. If this period is interpreted as the period of the global one-armed oscillation (GOAO), i.e., as distortions of the circumstellar disk in the form of a one-armed density wave, then it turns out to be the shortest among the known periods for Be stars.

The infrared observations in 1992–1995 [31] show that, first, not only the long-term but also the fairly rapid (of the order of a month) infrared variability correlates well with the optical variability. Second, the density in the circumstellar disk was estimated from the intensity of Paschen and Brackett emission lines (Pa11–Pa20 and Br11–Br20; the transition between the optically thin and optically thick radiation corresponds to Pa13–Pa15 and Br13–Br17) to be $N_e = (0.15–1.5) \times 10^{12}$ cm$^{-3}$. This value is in good agreement with our estimate $N_e = 10^{10}–10^{12}$ for a brightness lower than $V = 9\overset{m}{.}1$–$9\overset{m}{.}2$ (see Fig. 4); in 1992–1995, the brightness declined, on the average, from $9\overset{m}{.}0$ to $9\overset{m}{.}4$.

The photometric behavior of the star closely matches the behavior of a Be star during envelope ejection [6]. The pseudophotospheric temperature may decrease by a factor of 3 or 4, while the radius may increase by one order of magnitude or more. In our case, the B0III star has a radius of $\sim 15 R_\odot$ and $T_e = 28\,000$ K for $m_V = 9\overset{m}{.}50$ (at minimum). The colors at this minimum correspond to a spectral type slightly earlier than B0 (recall that the object's spectral type is O9.7IIIe). However, in 1953–1956, twenty years before the outburst, the colors corresponded to a spectral type slightly later than B0 (see Figs. 4 and 5); i.e., the stellar temperature turned out to be higher after the outburst when the shell phase ended. The star apparently "prepared" for the envelope ejection in the 1950s.

At maximum light m, $m_V = 8\overset{m}{.}85$, but the colors correspond to a blackbody with $T_e \sim 12\,000$ K. In order not only to offset the decrease in temperature but also to ensure the increase in luminosity, the radius of the emit-ting surface must increase tenfold. Thus, at maximum light we see not the B0 star but its cooler pseudophotosphere with an effective radius of $\sim 150 R_\odot$. This is probably the reason why there is a confusion with the distance determination for the object. For example, assuming that $M_V = -4.9$ (B0III) and $m_V = 8.9$, Wade and Oke [19] give a distance of 1.8 kpc. The same value was obtained by Giangrande et al. [3] for the spectral type O9.7IIIe and $m_V = 8.85$. Margon et al. [18] obtained a distance of 3.6 kpc by assuming that $M_V = -6$ (supergiant), while Janot-Pacheko et al. [32] give 2.6 kpc. The largest distance of 4.8 kpc was obtained by Stier and Liller [16]. In all cases, the visual absorption $A_V$ was assumed to be approximately the same, but all authors made the same mistake: the brightness at maximum was assumed to be that of the B0 star. In reality, the brightness at maximum is attributable to the emission from the circumstellar envelope with a much lower temperature, and only at minimum ($m_V = 9.50$) does the apparent magnitude correspond to the B0 star. Taking this into account, we obtained a distance of 2.9 kpc by assuming that the luminosity is $M_V = -5.0$ for the currently firmly established spectral type B0III. But the X-ray luminosity then turns out to be appreciably higher: $L_x = (1.5–2) \times 10^{-37}$ erg s$^{-1}$ for giant outbursts at a distance of 1.8 kpc and is a factor of 2.5 higher at $D = 2.9$ kpc.

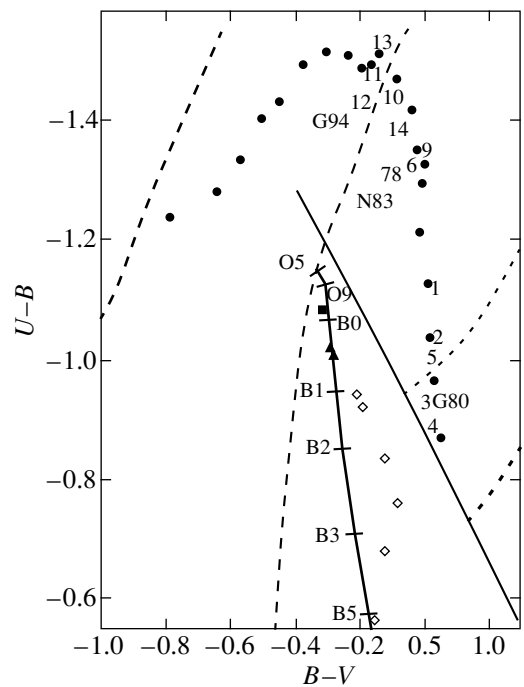

**Fig. 5.** Color variations (the numbers from 1 to 14) of the additional radiation during the giant X-ray outburst of 1980 (G80), the normal outburst of 1983 (N83), and the giant outburst of 1994 (G94). Numbers 3, 7, 8, and 12 correspond to the outburst maxima. The open diamonds give the positions of the mean color indices for the AD in HZ Her/Her X-1 for various fractions of the eclipsed AD part [30]; the upper diamond corresponds to half the AD observed at the beginning of the X-ray outburst. The remaining notation is the same as that in Fig. 4.

## CONCLUSION

Here, we present the *UBV* observations of the X-ray Be/XR binary system A0535+26/V725 Tau after 1988 and analyze all photoelectric *UBV* observations of the object (1973–1998) and its photographic observations since 1898. The X-ray activity is associated with the outburst of the Be star that began after 1970 and that is attributable to envelope ejection. The optical *V* luminosity reached a maximum in 1975, when the largest (in amplitude and duration) X-ray outburst was observed. The envelope dispersed almost completely in the fall of 1998; the colors at the minimum of 1998 (*minimum minimorum*) correspond to a spectral type slightly earlier than B0, while the colors in 1953–1956 (before the outburst) correspond to a slightly later spectral type than B0; at the same color excesses, i.e., after the outburst (the completion of the shell phase), the star became slightly hotter. All the additional (relative to the Be star) radiation is attributable to a single source, the circumstellar envelope (disk), and the contribution of the AD around the neutron star does not exceed a few percent.

The star appears to have passed all activity phases of Be stars: B–Be–Be shell–B. The Hα line was in absorption in the fall of 1998, the maximum intensity of the emission line was observed at the maximum light in 1975–1980, and the intensity of the Hα emission line decreased with declining brightness. The star has been known as a definitely emission-line star since 1942 [33], while in 1922 and 1926, it was classified as B0 [16].

In general, the optical behavior of the star is consistent with the existing theory of Be stars, except for the appearance of quasi-periodic light fluctuations ($P \approx 1400^d$) with an amplitude $\Delta V \approx 0.^m3$ after 1980, when the second giant X-ray outburst was observed. Since the circumstellar envelope dispersed almost completely in the fall of 1998, it would be unreasonable to expect strong X-ray outbursts in the coming years. The object will most likely be a weak X-ray source, as it was before 1975 [34]. The recurrence period of the envelope ejection in A0535+26/V725 Tau is not known. However, since only one such episode ~25 years in duration was observed in the 20th century, the next episode will not occur soon.

## ACKNOWLEDGMENTS

We wish to thank A.A. Aslanov for the 1980 observations and A.E. Tarasov, who provided the spectroscopic data of 1998.